\documentclass{aastex}

\begin{document}

\title{Interplanetary Network Localization of GRB991208 and the
Discovery of its Afterglow}
\author{K. Hurley}
\affil{University of California, Berkeley, Space Sciences Laboratory,
Berkeley, CA 94720-7450}
\email{khurley@sunspot.ssl.berkeley.edu}
\author{T. Cline}
\affil{NASA Goddard Space Flight Center, Code 661, Greenbelt, MD 20771}
\author{E. Mazets, R. Aptekar, S. Golenetskii, D. Frederiks}
\affil{Ioffe Physico-Technical Institute, St. Petersburg, 194021 Russia}
\author{D. Frail}
\affil{National Radio Astronomy Observatory, PO Box O, Socorro NM 87801}
\author{S. Kulkarni}
\affil{Palomar Observatory, 105-24, Caltech, Pasadena, CA 91125}
\author{J. Trombka, T. McClanahan}
\affil{NASA Goddard Space Flight Center, Code 691, Greenbelt, MD 20771}
\author{R. Starr}
\affil{The Catholic University of America, Department of Physics, Washington
DC 20064}
\author{J. Goldsten}
\affil{The Johns Hopkins University, Applied Physics Laboratory, Laurel, MD 20723}

\begin{abstract}
The extremely energetic ($\rm \sim 10^{-4} erg/cm^2$) gamma--ray burst (GRB) of 
1999 December 8 was 
triangulated to a $\sim$ 14 sq. arcmin.
error box $\sim$ 1.8 d after its arrival at Earth 
with the 3rd interplanetary network (IPN), consisting of the \it Ulysses, 
Near Earth Asteroid Rendezvous \rm (NEAR), 
and \it Wind \rm spacecraft.  Radio observations with the Very Large Array 
$\sim$ 2.7 d after the burst revealed a bright fading counterpart whose 
position is consistent
with that of an optical transient source whose redshift is z=0.707.
We present
the time history, peak flux, fluence, and refined 1.3 sq. arcmin. error box of this event, and discuss
its energetics.  This is the first time that a counterpart has been found
for a GRB localized only by the IPN.

\end{abstract}


\section{Introduction}

Many gamma-ray burst counterparts have
now been identified using the rapid, precise localizations available from
the BeppoSAX spacecraft ,
as well as from the \it Rossi X-Ray Timing Explorer \rm, starting with
GRB970228 (Costa et al. 1997; van Paradijs et al. 1997).  
Such detections occur at a low rate ($\sim 8 \, y^{-1}$), and they have been
limited to the long--duration events so far, 
but they have confirmed the
cosmological origin of at least this class of bursts.
Since 1977, interplanetary networks of omnidirectional GRB detectors have
provided precise triangulations of both short and long bursts at rates up to 
$\sim$ 1/week, but often
the networks have been incomplete, or the data return from the interplanetary
spacecraft has been slow.  The present, 3rd IPN is now complete with \it Ulysses \rm
and NEAR as its distant points (Cline et al. 1999) 
and, in conjunction with numerous near-Earth
spacecraft, can produce precise GRB error boxes within $\sim$ 1 d, making
them useful for multi-wavelength follow-up observations.  Here we present
the observations of GRB991208, which was rapidly localized to a small
error box, leading to the identification of its radio and optical afterglow, and
eventually to the measurement of its redshift.    

\section{IPN Observations}

GRB 991208 was observed by the \it Ulysses \rm GRB (Hurley et al. 1992), KONUS-\it Wind \rm 
(Aptekar et al. 1995), and NEAR X-ray/Gamma-ray Spectrometer (XGRS: Goldsten et al.
1997) experiments.   \it Ulysses \rm, in heliocentric orbit, NEAR, approaching rendezvous 
with the asteroid Eros, and \it Wind \rm, were 2176, 937, and 1.5 light-seconds
from Earth, respectively.  

We focus here on the \it Ulysses \rm and NEAR data; KONUS data
will be presented elsewhere.
The \it Ulysses \rm and XGRS light curves are shown in figure 1.  
Although \it Ulysses \rm recorded the first 57 s of the burst with
0.03125 s resolution in the triggered mode, the peak of the event occurred slightly later,
and we have shown the 0.5 s resolution real-time data in the figure.
The XGRS
BGO anticoincidence shield is employed as the NEAR burst monitor, and the only
time history data available from it are 1 s resolution count rates for the 100 - 1000 keV energy
range.  The burst
can be characterized by a T$_{90}$ duration of 68 s, placing
it firmly in the ``long'' class of bursts (Hurley 1992; Kouveliotou et al.
1993).   The event-integrated \it Ulysses \rm spectrum is well fit between 25 and 150 keV
by a power law, thermal bremsstrahlung, or blackbody spectrum.  For the following
analysis, we adopt the power law, which has a photon index 1.68 $\pm$ 0.19,
and a $\chi^2$ of 4.3 for 11 degrees of freedom (figure 2).
The 25 - 100 keV fluence is $\rm 4.9 \times 10^{-5} \,erg \,cm^{-2}$, with an
uncertainty of $\sim \rm \pm 10\% $ due to count rate statistics and systematics.  Since
the XGRS shield provides only very low resolution (40 minutes) spectral data, we can
only estimate the fluence in the XGRS energy range from its light curve to be about the
equivalent of that observed by \it Ulysses \rm.  Thus the total fluence above 25 keV
is $\sim \rm 10^{-4}\, erg\, cm^{-2}$.  Bursts with this intensity or 
greater occur at a rate $\rm \lesssim 10 \,y^{-1}$.  Since the peak of the event occurs when
only the 0.5 s \it Ulysses \rm data are available, we can estimate the peak
flux only over this time interval; it is $\rm \sim 5.1 \times 10^{-6}\, erg\, cm^{-2}\, s^{-1}$,
25 - 100 keV ($\sim \pm \rm 10\% $), with a contribution in the 100 - 1000 keV energy range which is
again probably equivalent.   

A preliminary $\sim$ 14 
sq. arcmin. IPN error box was circulated $\sim$ 44 h after the earth-crossing time
of the event (04:36:53 UT: Hurley et al. 1999).  The final error box
is shown in figure 3, and nests within the preliminary one; it has an area of 
$\sim$ 1.3 sq. arcmin.  Its coordinates are given in Table 1.  
Also shown in figure 3 is the position of the radio counterpart to the
burst detected by Frail (1999).  We note that this is the first
time that a GRB position determined by the 3rd IPN alone (i.e., no \it
Compton Gamma-Ray Observatory \rm BATSE or \it BeppoSAX \rm observations)
has been successfully
used for multiwavelength counterpart searches. 

\section{Observations with NRAO Very Large Array}

Very Large Array (VLA)\footnotemark\footnotetext{The NRAO is a facility of the National
Science Foundation operated under cooperative agreement by Associated
Universities, Inc.} observations were begun on 1999 December 10.92 UT, 2.73 days
after the gamma-ray burst. In order to image the entire 14 sq. arcmin. initial
IPN error box (Hurley et al. 1999) with the VLA at 8.46 GHz two
pointings were required, each with a field-of-view at half power
response of 5.3\arcmin. The full 100 MHz bandwidth was used in two
adjacent 50-MHz bands centered at 8.46 GHz. A single pointing was also
made at 4.86 GHz, but the bandwidth was halved in order to image the full
9\arcmin \,field-of-view without distortion.  The flux
density scale was tied to the extragalactic sources 3C\thinspace{48}
(J0137+331), while the array phase was monitored by switching between
the GRB and the phase calibrators J1637+462 (at 8.46 GHz) and J1658+476
(at 4.86 GHz). 

There were three radio sources inside the initial IPN error box (figure 3). Of these, two
were previously known from an earlier survey of this part of the sky
(Becker, White \& Helfand 1995). The third source, located at located at
(epoch J2000) $\alpha$\ =\ $16^h33^m53.50^s$ ($\pm{0.01^s}$) $\delta$\
=\ $+46^\circ27^\prime20.9^{\prime\prime}$ ($\pm{0.1}^{\prime\prime}$)
was near the center of the initial IPN error box.  On the basis of its position,
compactness ($<0.8^{\prime\prime}$), and a rising flux density between
4.86 GHz and 8.46 GHz (327 $\pm$ 45 $\mu$J and 707 $\pm$ 39 $\mu$J respectively)
, Frail (1999) proposed that it was the afterglow
of GRB\thinspace{991208}.  Despite the proximity of this location to the Sun
($\rm \sim 70 \arcdeg$), numerous optical observations were carried out,
and this suggestion was quickly confirmed by the
independent detection of a coincident optical source, not visible on the
Digital Sky Survey (Castro-Tirado et al. 1999). 

In the week following the afterglow discovery, it was determined that
the optical flux faded as a power-law with a rather steep temporal decay
index $\alpha\simeq 2.15$ (where $F_\nu\propto t^\alpha$) (Jensen et al.
1999; Garnavich and Noriega-Crespo 1999; Masetti et al.
1999). Optical spectroscopy emission lines from a presumed
host galaxy, if identified with [OII] and [OIII] features, place
GRB\thinspace{991208} at a redshift $z=0.707 \pm 0.002$ (Dodonov et al
1999), making this the third closest GRB with a spectroscopically-measured
redshift. GRB\thinspace{991208} has been no less interesting at
radio wavelengths.  It has the brightest radio afterglow detected to
date and consequently it has been detected and is well-studied between 1
GHz and 350 GHz (Pooley 1999; Shepherd et al. 1999; Bremer et al. 1999).

\section{Discussion}

For a 25 -- 1000 keV fluence of $\rm 10^{-4}\, erg\, cm^{-2}$ and
a redshift z=0.707, the isotropic energy of this burst would have
been $\rm 1.3 \times 10^{53} erg$; for a peak flux of $\rm \sim 10^{-5}\, erg\, cm^{-2}\, s^{-1}$
in the same energy range, the isotropic peak luminosity would have been
$\rm 1.3 \times 10^{52} erg\, s^{-1}$.  These estimates assume $\rm \Omega=0.2,
\Lambda=0, and \, H_0=65 \, km\, s^{-1}\, Mpc^{-1}$.  Rhoads (1997) has
pointed out that one signature of beaming is a steep decay in the
afterglow light curve, $\propto t^{-2}$.  As the initial optical light
curve for GRB991208 indeed appears to decay this steeply, the emission
may well be beamed, reducing these estimates.

The current IPN now consists of \it Ulysses \rm and NEAR in interplanetary
space, and numerous near-Earth spacecraft such as \it Wind \rm, \it BeppoSAX \rm, and the
\it Compton Gamma-Ray Observatory \rm.  The \it Mars Surveyor 2001 Orbiter \rm will
join the network in mid-2001.  The IPN currently observes $\sim$ 1 -- 2 GRBs per
week and is localizing many of them rapidly to small error boxes.  These
events tend to be the brighter ones, but apart from this, there is no
bias towards any particular event duration; indeed, the IPN generally
obtains its smallest error boxes for the short bursts.  Neither is there
any sun-angle restriction for the event locations, which means that bursts
will be detected whose locations are close to the Sun, as this one was,
making prompt radio
observations of these positions important.  The other advantages of
radio observations over optical are the longer lifetime of the radio afterglow,
the immunity from weather, and the freedom to operate at any part of the
diurnal cycle.
This should increase the rate of counterpart detections substantially over the next
several years.

KH acknowledges support for Ulysses operations under JPL Contract 958056,
for IPN operations under NASA LTSA grant NAG5-3500, and for NEAR operations
under the NEAR Participating Scientist program.  On the
Russian side, this work was partially supported by RFBR grant \# 99-02-17031.  
We are grateful to R. Gold and R. McNutt for their
assistance with the NEAR spacecraft.  RS is supported by NASA grant
NCC5-380.  We are indebted to T. Sheets for her excellent work on NEAR
data reduction.  Special thanks also go to the NEAR project office for its support of
post-launch changes to XGRS software that made these measurements possible.  In
particular, we are grateful to John R. Hayes and Susan E. Schneider for writing
the GRB software for the XGRS instrument and to Stanley B. Cooper and David S. Tillman
for making it possible to get accurate universal time for the NEAR GRB detections.

\clearpage

\begin{figure}
\figurenum{1}
\epsscale{.75}
\plotone{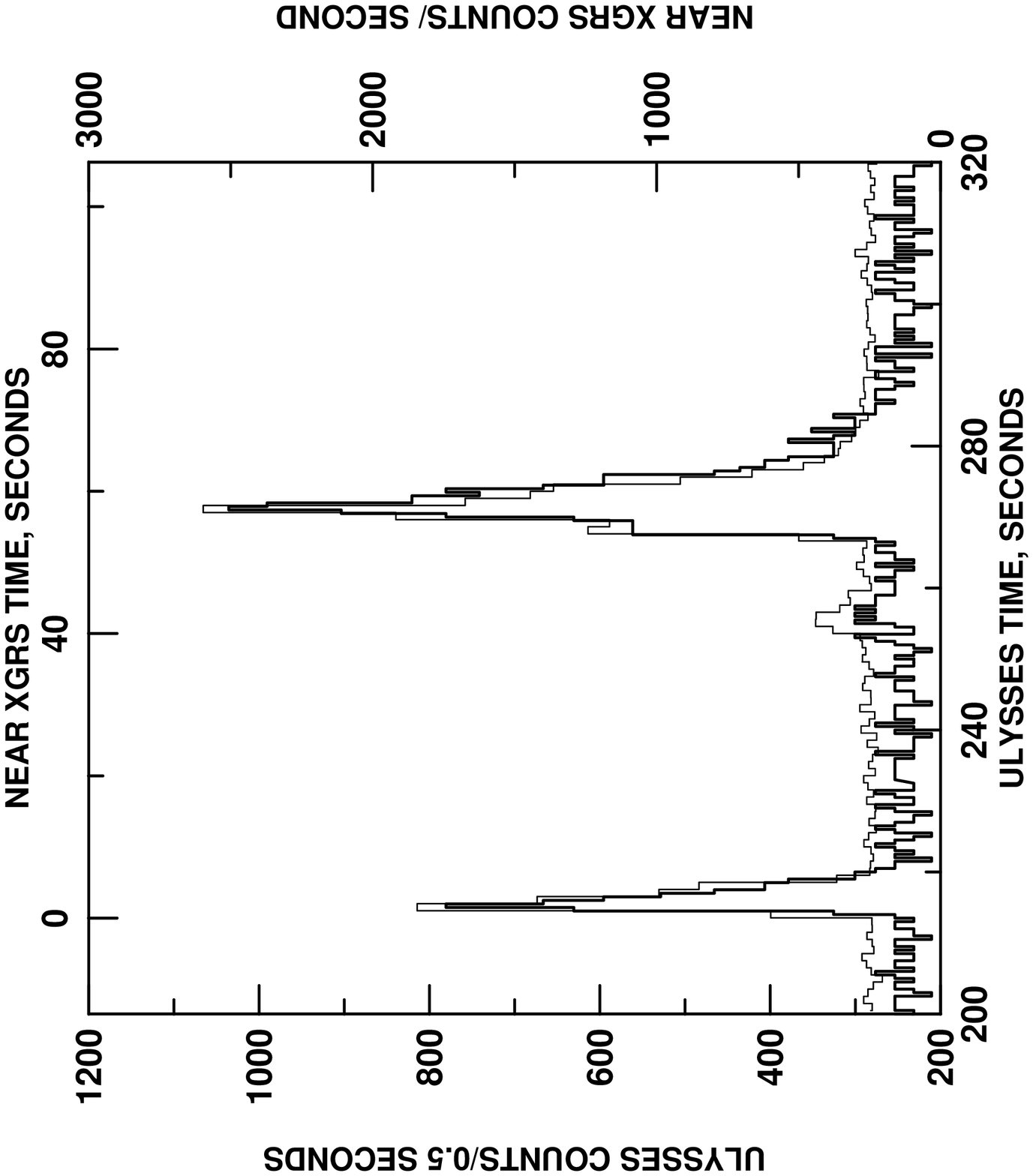}
\caption{Time histories of GRB991208 as observed by \it Ulysses \rm GRB (22-150 keV, thick line) and
NEAR XGRS (100-1000 keV, thin line).  The \it Ulysses \rm data are compressed onboard
the spacecraft and decompressed on the ground, leading to discrete count rate levels.}
\end{figure}

\begin{figure}
\figurenum{2}
\epsscale{.75}
\plotone{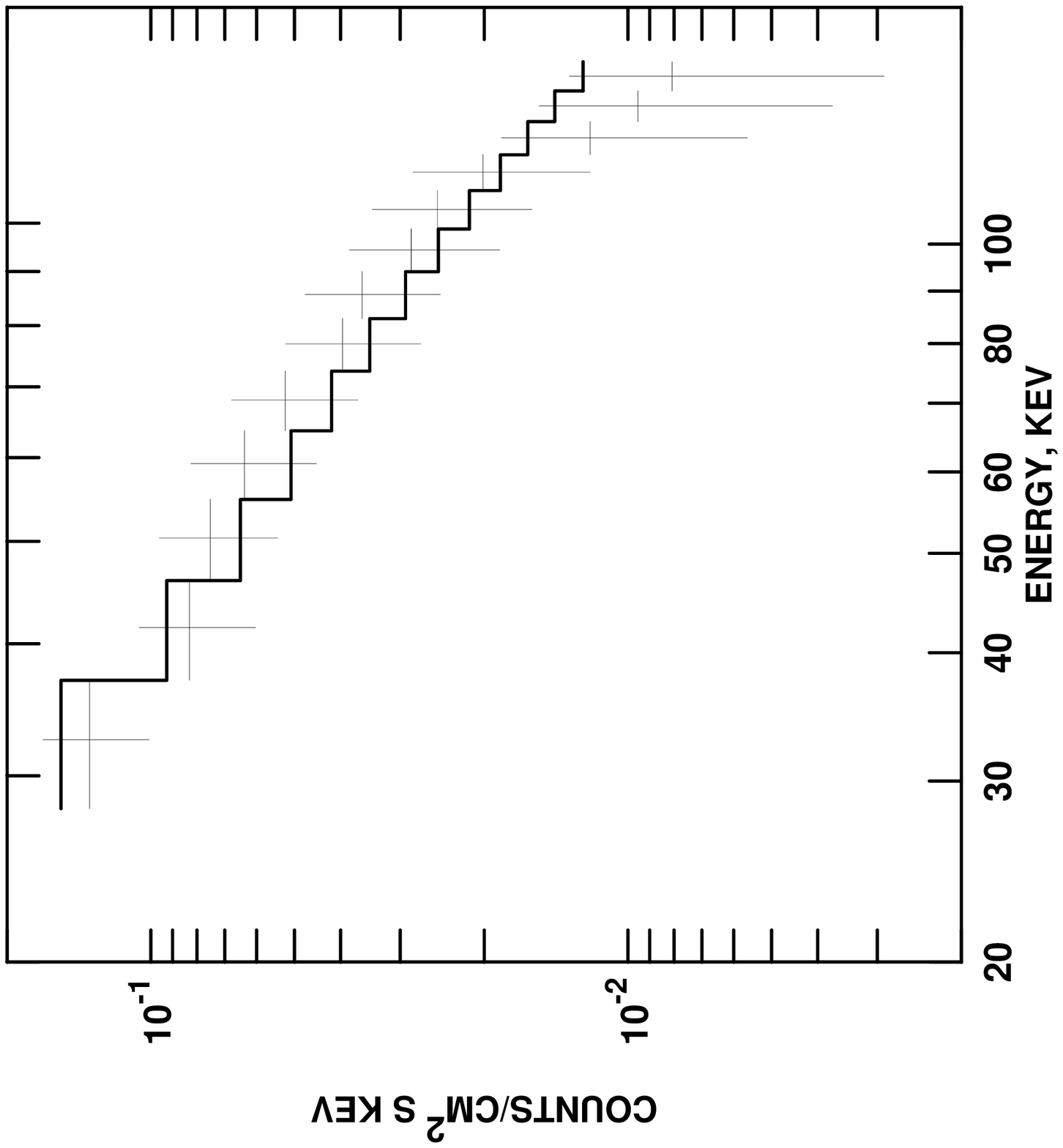}
\caption{Energy spectrum of GRB991208 as observed by Ulysses GRB (crosses) and
the fitted power law spectrum (solid line).}
\end{figure}

\begin{figure}
\figurenum{3}
\epsscale{}
\plotone{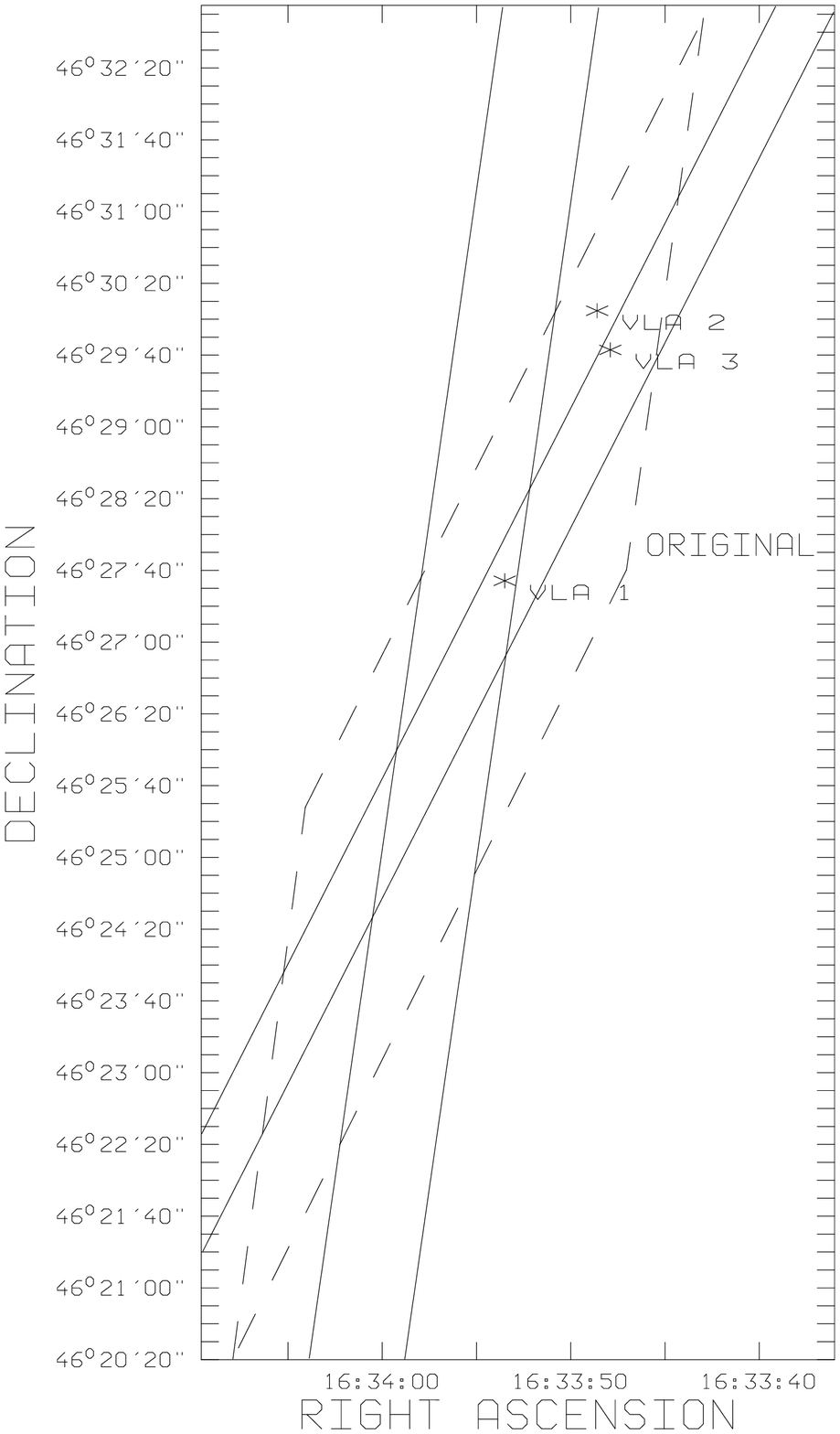}
\caption{Original and final IPN error boxes for GRB991208 (3 $\sigma$ confidence).
The original error box (Hurley et al. 1999) is drawn with dashed lines.  The final
error box is defined by the intersection of the 30 \arcsec \it Ulysses \rm-KONUS annulus
and the 52 \arcsec \it Ulysses \rm-NEAR annulus, and has an area of 1.3 sq. arcmin.
Three sources are indicated.  Two field sources in the upper right-hand corner are
marked VLA 2 and VLA 3; their flux densities are 3.4 and 0.4 mJy respectively.  VLA
1 is the counterpart.  It lies at roughly at the 96 \% confidence contour.  }
\end{figure}

\clearpage

\begin{deluxetable}{ccc}
\tablecaption{IPN error box for GRB991208 (3 $\sigma$ confidence).}
\tablehead{
\colhead{} & \colhead{$\rm \alpha (2000)$, degrees} & \colhead{$\rm \delta(2000)$, degrees}
}
\startdata
Center:     & 248.4848        &  46.4413 \\
Corners:    & 248.4727        &  46.4487 \\
            & 248.5021        &  46.4078 \\
            & 248.4674        &  46.4747 \\
            & 248.4969        &  46.4338 \\

\enddata

\end{deluxetable}

\end{document}